\documentclass[a4paper]{jpconf}
\usepackage{graphicx}
\usepackage{hyperref}
\begin{document}
\title{Excitation spectra of solitary waves in scalar field models with polynomial self-interaction}

\author{Vakhid A. Gani$^{1,2}$, Vadim Lensky$^{1,2,3}$, Mariya A. Lizunova$^{1,2}$, Elizaveta V. Mrozovskaya$^1$}

\address{$^1$National Research Nuclear University MEPhI (Moscow Engineering Physics Institute), Kashirskoe highway 31, Moscow, 115409, Russia}
\address{$^2$National Research Centre Kurchatov Institute, Institute for Theoretical and Experimental Physics, Bolshaya Cheremushkinskaya str. 25, Moscow, 117218, Russia}
\address{$^3$Institut f\"ur Kernphysik \& PRISMA  Cluster of Excellence, Johannes Gutenberg Universit\"at Mainz,  D-55128 Mainz, Germany}

\ead{vakhid.gani@gmail.com}

\begin{abstract}
We study excitations of solitary waves -- the kinks -- in scalar models with degree eight polynomial self-interaction in $(1+1)$ dimensions. We perform numerical studies of scattering of two kinks with an exponential asymptotic off each other and analyse the occurring resonance phenomena. We connect these phenomena to the energy exchange between the translational and the vibrational modes of the colliding kinks. We also point out that the interaction of two kinks with power-law asymptotic can lead to a long-range interaction between the two kinks.
\end{abstract}

\section{Introduction}

Field-theoretical models with polynomial self-interaction are widespread in modern theoretical physics, being applied to various phenomena from cosmology to condensed matter physics. In the latter, the most known example is perhaps the phenomenological Ginzburg-Landau model, where the complex scalar field of Cooper pairs has a polynomial self-interaction of the fourth degree. The $\varphi^4$ model with two degenerated minima is commonly used as a model example of spontaneous symmetry breaking, whereas models with more sophisticated potentials of higher polynomial degrees are applied, e.g., to model several consecutive phase transitions~\cite{khare,pavlov}.

We further consider field models with a single real field and polynomial potentials that are even functions of field strength $\varphi$ and have degenerate minima. Such models admit topological solutions --- the kinks --- that interpolate between two adjacent minima of the potential (the vacua of the model). The kinks of the $\varphi^4$ and $\varphi^6$ models are well studied, yielding many interesting results concerning, e.g., excitations of the kinks~\cite{saad02,saad03} and scattering of two kinks off each other~\cite{dorey01,GaKuLi}. At the same time, similar features of models with self-interaction of higher degrees, used in studies of many important phenomena~\cite{poltis1}, have only recently started to be explored systematically~\cite{khare,GaLeLi}.

Our work studies excitations of the $\varphi^8$ kinks and their relation to resonance phenomena in kink-antikink collisions. We connect the occurrence of so-called escape windows to the presence of vibrational excitations.

\section{A kink of the $\varphi^8$ model and its excitation spectrum}

The $\varphi^8$ model that we consider is described by the following Lagrangian in $(1+1)$ space-time dimensions, with $\varphi=\varphi(t,x)$ being a real scalar field:
\begin{equation}
{\cal L}=\frac{1}{2} \left(\frac{\partial\varphi}{\partial t} \right)^2-\frac{1}{2} \left(\frac{\partial\varphi}{\partial x} \right)^2-V(\varphi),	
	\label{eq:largang}
\end{equation}
where the potential $V(\varphi)$ is a degree eight polynomial with two, three, or four degenerate minima $\varphi_i$ yielding $V(\varphi_i)=0$. This Lagrangian develops topological BPS-saturated solutions $\varphi_{i_1 i_2}(x)$, each of them connecting two adjacent minima, $\varphi_{i_1}$ and $\varphi_{i_2}$~\cite{manton}. To explore the excitation spectra of the kinks, one starts along the standard procedure used to study the linear Lyapunov stability: the kink is perturbed by a small deviation,
\begin{equation}
    \delta\varphi(t,x)=\psi(x)\cos(\omega t),
    \nonumber
    \label{eq:subexpan}
\end{equation}
and the spectrum of $\omega^2$ coincides to that of the time-independent Schr\"odinger-like
problem with the potential
\begin{equation}
    U(x)=\left.\frac{d^2V}{d\varphi^2}\right|_{\varphi_{\scriptsize \mbox{k}}(x)}.
    \label{eq:uxexpan}
    \nonumber
\end{equation}
$U(x)$ is a potential well with different or identical asymptotic values at $x\to\pm\infty$ depending on the symmetry of the initial kink. The BPS kinks are classically stable, hence $\omega^2$ cannot be negative; apart from that, it is easy to show that there always is an eigenmode with $\omega^2=0$ --- the translational (zero) mode. There can also be one or more discrete (vibrational) modes with $\omega^2>0$, lying below the continuum part of the spectrum.

We further consider a variant of the $\varphi^8$ that has four degenerate vacua, corresponding to the self-interaction potential
\begin{equation}
    V(\varphi)=(\varphi^2-a^2)^2(\varphi^2-b^2)^2
    \label{eq:potphi8}
\end{equation}
that has four degenerate minima $\pm a$, $\pm b$. Following~\cite{khare}, we set $a=(\sqrt{3}-1)/2$, $b=(\sqrt{3}+1)/2$.
Equation (\ref{eq:potphi8}) admits three topological sectors, $(-b,-a)$, $(-a,a)$, and $(a,b)$, each with its kink and the corresponding antikink. We constrain our study to the sector $(-a,a)$, with the kink profile given by~\cite{khare,lohe}:
\begin{equation}
    e^{\mu x}=\frac{a+\varphi}{a-\varphi}\left(\frac{b-\varphi}{b+\varphi}\right)^{a/b},
    \label{eq:kinkph8a}
\end{equation}
where $\mu=2\sqrt{2}a(b^2-a^2)$. The antikink of this sector is obtained from equation (\ref{eq:kinkph8a}) by substituting $x\to -x$. 

Our numerical study of the excitation spectrum of this kink confirms the existence of the zero mode (our result is $\omega^2_0=-2\times 10^{-8}$, the deviation of which from zero can serve as an uncertainty estimate), and also finds a vibrational mode corresponding to $\omega^2_1=2.70491$. The latter excitation manifests itself in resonance phenomena in kink-antikink collsions, as explained below.

\section{Kink-antikink collisions in the $\varphi^8$ model}

A vibrational mode in the excitation spectrum of a kink (antikink) leads under certain conditions to a part of the kink's kinetic energy being transformed into small oscillations of the kink's profile --- the resonant energy exchange between the translational and the vibrational modes. This gives rise to a host of exciting phenomena in kink-antikink collsions~\cite{dorey01,GaKuLi,aek01,GaKuPRE}.

In order to explore how this mechanism works in the $\varphi^8$ model, we performed a numerical study of kink-antikink scattering in the topological sector described in the previous section, with the self-interaction given by equation (\ref{eq:potphi8}). The initial configuration was taken as the sum of the kink (\ref{eq:kinkph8a}) and the corresponding antikink, located at $t=0$, respectively, at $x=\pm x_0$, $x_0=12.5$, and moving towards each other with the respecitve initial velocities $\pm v_{\scriptsize \mbox{in}}$.

\begin{figure}[h]
\textrm{\small\it (a)}
\begin{minipage}[h]{0.29\linewidth}
\center{\includegraphics[width=0.9\linewidth]{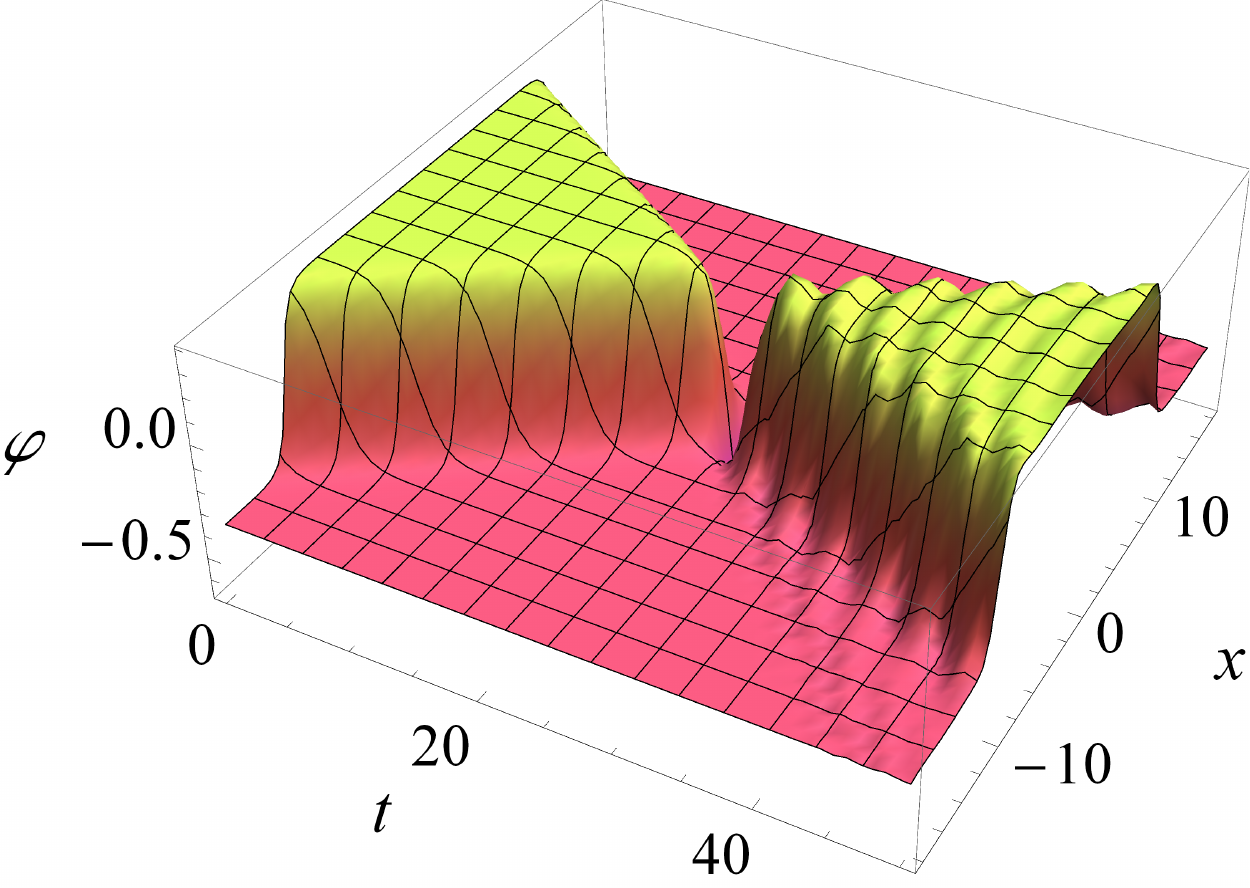}}
\end{minipage}
\textrm{\small\it (b)}
\begin{minipage}[h]{0.29\linewidth}
\center{\includegraphics[width=0.9\linewidth]{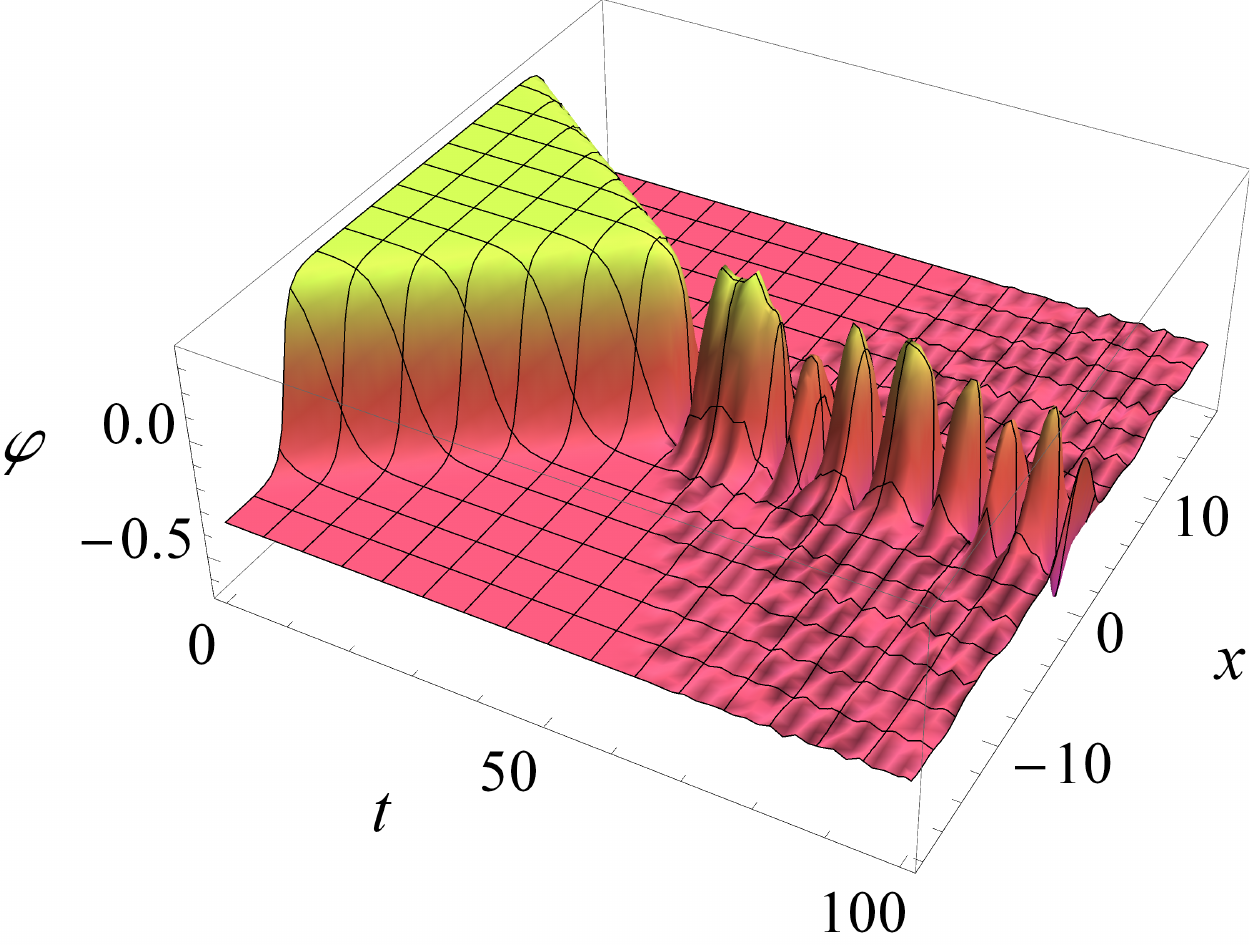}}
\end{minipage}
\textrm{\small\it (c)}
\begin{minipage}[h]{0.29\linewidth}
\center{\includegraphics[width=0.9\linewidth]{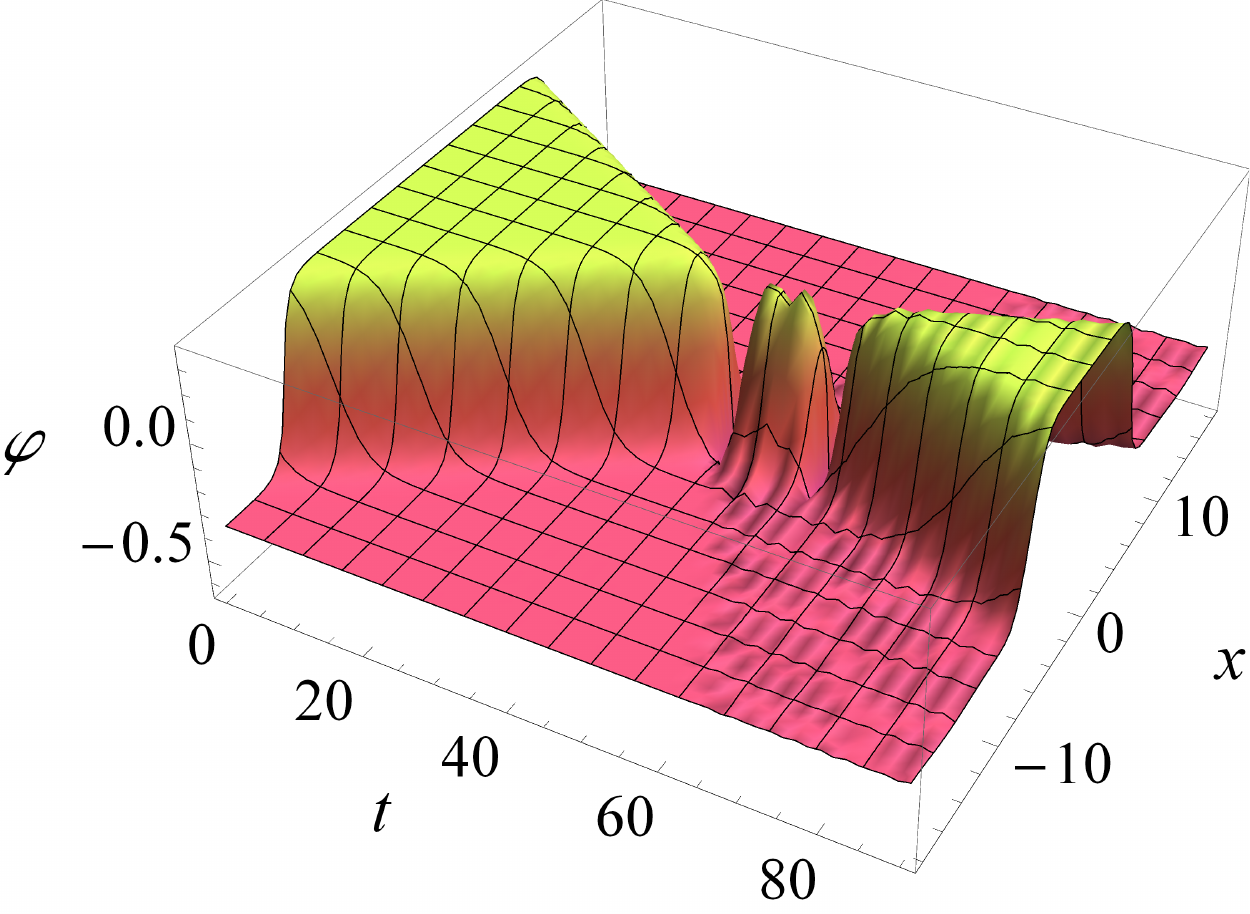}}
\end{minipage}
\caption{a) Escape at $v_\mathrm{in}>v_\mathrm{cr}$; b) Formation of a bion at $v_\mathrm{in}<v_\mathrm{cr}$;
c) Escape window: escape after two bounces at $v_\mathrm{in}<v_\mathrm{cr}$.}
\label{fig:kink4}
\end{figure}

We found that there are several different scattering regimes, depending on the value of $v_\mathrm{in}$. In particular, there is a critical value $v_\mathrm{in}=v_{\scriptsize \mbox{cr}}\simeq 0.3160$ such that the kink and the antikink bounce off each other at all $v_{\scriptsize \mbox{in}}\ge v_{\scriptsize \mbox{cr}}$, see figure \ref{fig:kink4}a. This process is only approximately elastic, i.e., $v_{\scriptsize \mbox{f}}<v_{\scriptsize \mbox{in}}$, which can also be seen in figure \ref{fig:kink4}. For example, $v_\mathrm{f}=0.23$ at $v_\mathrm{in}=0.40$, $v_\mathrm{f}=0.36$ at $v_\mathrm{in}=0.50$, and $v_\mathrm{f}=0.47$ at $v_\mathrm{in}=0.60$. At $v_{\scriptsize \mbox{in}}<v_{\scriptsize \mbox{cr}}$, on the other hand, the two kinks capture each other and form a long-lived quasi-bound state, the bion, see figure \ref{fig:kink4}b. This, however, does not happen universally at all $v_{\scriptsize \mbox{in}}<v_{\scriptsize \mbox{cr}}$: we found narrow intervals of $v_\mathrm{in}$ --- the escape windows --- where the kink and the antikink escape after two, three, or more bounces (see figure \ref{fig:kink4}c, which shows an escape after two bounces). The phenomenon of escape windows is explained by the resonance energy exchange between the translational and the vibrational modes of the kink (antikink). The resonance condition
\begin{equation}
  \omega_R T_{12}=2\pi n+\delta,
  \label{eq:rescond}
\end{equation}
where $T_{12}$ is the time between the two bounces, $\delta$ is a constant,
and $n$ is an integer number, resulted in the empirical value of the resonance frequency $\omega_R = 1.618$ being very close to the frequency of the kink's vibrational
mode, $\omega_1=\sqrt{\omega_1^2}=1.644$.

\section{Conclusion}

In conclusion, we performed a first study of excitation spectra of the kinks occurring in the $\varphi^8$ field model. We connected the occurrence of a vibrational excitation of a single kink to the resonant energy exchange between this excitation and the translational excitation of the kink, leading to escape windows in the kink-antikink collisions.

We would like to mention a case of special interest that can be realized in various field models. If the model has a vacuum where the second derivative of the self interaction potential turns to zero, the kink that approaches this vacuum does so under a power-law asymptotic, in distinction to the situation considered previously where the asymptotic is exponential. Such kinks can occur in the $\varphi^8$  model or models with higher polynomial degrees under certain choice of the potential parameters~\cite{khare,lohe}. Qualitatively, such a power-law asymptotic leads to a long-range interaction between the kink and the corresponding antikink. In particular, it would be interesting to study how well a kink-antikink system with long-range interaction can be treated using the collective coordinate method; we plan to perform such a study of the corresponding sector of the $\varphi^8$ model in the future.

\section*{Acknowledgments}

The authors are very grateful to Prof.~A.~E.~Kudryavtsev for his interest to their work and for enlightening discussions. This work was supported in part by the Russian Federation Government under grant No.~NSh-3830.2014.2. V.~A.~Gani acknowledges support of the Ministry of Education and Science of the Russian Federation, Project No.~3.472.2014/K. M.~A.~Lizunova thanks the ITEP support grant for junior researchers and gratefully acknowledges financial support from the Dynasty Foundation.

This work was performed within the framework of the Center of Fundamental Research and Particle Physics supported by MEPhI Academic Excellence Project (contract No.~02.a03.21.0005, 27.08.2013).


\section*{References}
\begin{thebibliography}{9}

\bibitem{khare}
A.~Khare, I.~C.~Christov, and A.~Saxena, \href{http://dx.doi.org/10.1103/PhysRevE.90.023208}{Phys.~Rev.~E {\bf 90}, 023208 (2014)}; \href{http://arxiv.org/abs/1402.6766}{arXiv: 1402.6766 [math-ph]}.

\bibitem{pavlov}
S.~V.~Pavlov, M.~L.~Akimov, Crystall. Rep.~{\bf 44}, 297 (1999).

\bibitem{saad02}
D.~Saadatmand, S.~V.~Dmitriev, D.~I.~Borisov, P.~G.~Kevrekidis, M.~A.~Fatykhov, and K.~Javidan, \href{http://dx.doi.org/10.7868/S0370274X15070140}{Pis’ma v ZhETF {\bf 101}, 550 (2015)} [\href{http://dx.doi.org/10.1134/S0021364015070140}{JETP Lett.~{\bf 101}, 497 (2015)}].

\bibitem{saad03}
D.~Saadatmand, S.~V.~Dmitriev, D.~I.~Borisov, P.~G.~Kevrekidis, M.~A.~Fatykhov, and K.~Javidan, \href{http://dx.doi.org/10.1016/j.cnsns.2015.05.012}{Commun.~ Nonlinear Sci.~Numer.~Simulat.~{\bf 29}, 267 (2015)}; \href{http://arxiv.org/abs/1411.5857}{arXiv: 1411.5857 [nlin.PS]}.

\bibitem{dorey01}
P.~Dorey, K.~Mersh, T.~Romanczukiewicz, and Y.~Shnir,
\href{http://dx.doi.org/10.1103/PhysRevLett.107.091602}{Phys.~Rev.~Lett.~{\bf 107}, 091602 (2011)}; \href{http://arxiv.org/abs/1101.5951}{arXiv:1101.5951 [hep-th]}.

\bibitem{GaKuLi}
V.~A.~Gani, A.~E.~Kudryavtsev, and M.~A.~Lizunova,
\href{http://dx.doi.org/10.1103/PhysRevD.89.125009}{Phys.~Rev.~D {\bf 89}, 125009 (2014)}; \href{http://arxiv.org/abs/1402.5903}{arXiv:1402.5903 [hep-th]}.

\bibitem{poltis1}
E.~Greenwood, E.~Halstead, R.~Poltis, and D.~Stojkovic, \href{http://dx.doi.org/10.1103/PhysRevD.79.103003}{Phys.~Rev.~D {\bf 79}, 103003 (2009)}.

\bibitem{GaLeLi}
V.~A.~Gani, V.~Lensky, and M.~A.~Lizunova,
\href{http://dx.doi.org/10.1007/JHEP08(2015)147}{JHEP {\bf 08}, 147 (2015)}; \href{http://arxiv.org/abs/1506.02313}{arXiv: 1506.02313 [hep-th]}.

\bibitem{manton}
N.~Manton, P.~Sutcliffe, {\it Topological Solitons} (Cambridge University Press, Cambridge, 2004).

\bibitem{lohe}
M.~A.~Lohe, \href{http://dx.doi.org/10.1103/PhysRevD.20.3120}{Phys.~Rev.~D {\bf 20}, 3120 (1979)}.

\bibitem{aek01}
T.~I.~Belova and A.~E.~Kudryavtsev, \href{http://dx.doi.org/10.3367/UFNr.0167.199704b.0377}{Usp.~Fiz.~Nauk {\bf 167}, 377 (1997)} [\href{http://dx.doi.org/10.1070/PU1997v040n04ABEH000227}{Phys.-Usp.~{\bf 40}, 359 (1997)}].

\bibitem{GaKuPRE}
V.~A.~Gani and A.~E.~Kudryavtsev, \href{http://dx.doi.org/10.1103/PhysRevE.60.3305}{Phys.~Rev.~E {\bf 60}, 3305 (1999)}; \href{http://arxiv.org/abs/cond-mat/9809015}{arXiv: cond-mat/9809015}.
 
\end{thebibliography}
\end{document}